\documentclass[letter,twocolumn,A4]{jpsj3}
\usepackage{bm}

\title{Evidence for a Dual-Source Mechanism of THz Radiation from Rectangular Mesas of Single Crystalline \bm{$\mathrm{Bi_2Sr_2CaCu_2O_{8+\delta}}$} Intrinsic Josephson Junctions}

\author{Kazuo KADOWAKI, Manabu TSUJIMOTO, Kazuhiro YAMAKI, Takashi YAMAMOTO, Takanari KASHIWAGI, Hidetoshi MINAMI, Masashi TACHIKI and Richard A. KLEMM$^1$%\\
}
\inst{Graduate School of Pure \& Applied Sciences, University of Tsukuba, 1-1-1, Tennodai, Tsukuba, Ibaraki 305-8573, Japan\\
$^1$Department of Physics, University of Central Florida, Orlando, FL 32816, USA
}
\abst{The THz radiation emitted from rectangular mesas of single-crystal $\mathrm{Bi_2Sr_2CaCu_2O_{8+\delta}}$ was studied using angular distribution measurements and Fourier transform infrared spectroscopy.  Unlike the recent theoretical predictions, the results provide strong evidence for a dual-source mechanism in which the uniform and non-uniform parts of the $ac$-Josephson current act as electric and magnetic current sources, respectively.  The latter synchronizes with cavity modes of the mesa with integral harmonics of the fundamental radiation.\\
\\
\\
\noindent
Keywords: THz radiation, THz emitters, intrinsic Josephson junctions, Josephson plasma, $\mathrm{Bi_2Sr_2CaCu_2O_{8+\delta}}$ single crystals, $ac$-Josephson effect, high-$T_c$ superconductors, continuous THz sources\par}

\begin{document}
\maketitle
The recent discovery of coherent electromagnetic waves at terahertz frequencies (1 THz=$10^{12}$ Hz) from the intrinsic Josephson junctions (IJJs) within a mesa fabricated from high-quality single crystals of the high transition temperature $T_c$ superconductor $\mathrm{Bi_2Sr_2CaCu_2O_{8+\delta}}$ (BSCCO), presently denoted as STAR (Stimulated Terahertz Amplified Radiation) emitter, has triggered a great deal of research interest.  Besides fundamental studies of the physics and chemistry, there has been a great interest in the development of possible compact and all solid-state THz radiation sources, which may be very useful for applications in science and engineering fields such as medicine, diagnostics, pharmaceutical development, bioscience, ultrahigh-speed communication, environmental studies, security issues, and various types of nondestructive and noninvasive sensing and imaging, \textit{etc.}\cite{1}.   Some basic features of the STAR emitter were previously reported\cite{2,3}: the emission is stable, continuous and powerful ($\sim \mu$W), and its spectral width is less than the resolution limit of the Fourier transform infrared (FTIR) spectrometer, 7.5 GHz.  By varying the applied $dc$ voltage $V$ across the stack of $N$ IJJs, the frequency of the THz radiation appears to be locked by a cavity resonance mechanism, which fixes it to that of one of the standing wave modes inside the mesa.  However, the precise nature of the radiation mechanism as well as the fundamental features were not understood yet.  Moreover, to develop the STAR emitter with greater power for useful applications, it is first important to understand the physical nature of the emission mechanism experimentally.

Theoretically, a unique radio engineering approach to the mechanism was made by Pedersen and Madsen\cite{4}, whereas models based on numerical simulations of the coupled Sine-Gordon equations for stacked Josephson junctions were widely used by others\cite{5,6,7,8,9,10,11,12,13,14}.  Among them some models\cite{5,6,7,8,9,10} assumed a magnetic current radiation source (or cavity mode) only without taking into account the electromagnetic fields at near-field, and also neglecting the substrate effect.  Here we provide clear experimental evidence contradicting those model calculations.

We measured the angular dependence of the far-field radiation from various rectangular mesas.  Our experimental results are inconsistent with the predictions of either an electric or a magnetic surface current sources acting alone.  Instead, they strongly suggest a dual-source THz radiation mechanism\cite{15}:  the uniform part of the $ac$-Josephson current acts as an electric surface current source, and the inhomogeneous part of the $ac$-Josephson current sets up a displacement current that excites a cavity resonance mode in the mesa, which locks the radiation frequency and acts as a magnetic surface current source.  By adjusting the relative amplitude and phase of the two source currents and accounting for the substrate, excellent agreement with experiment is obtained.  Furthermore, from FTIR analysis of the $dc$ current-voltage (\textit{I-V}) characteristic, the locking onto the resonant cavity mode appears to occur by the sequential synchronization of each of the IJJs.

\begin{figure}[h]
\includegraphics[width=0.45\textwidth]{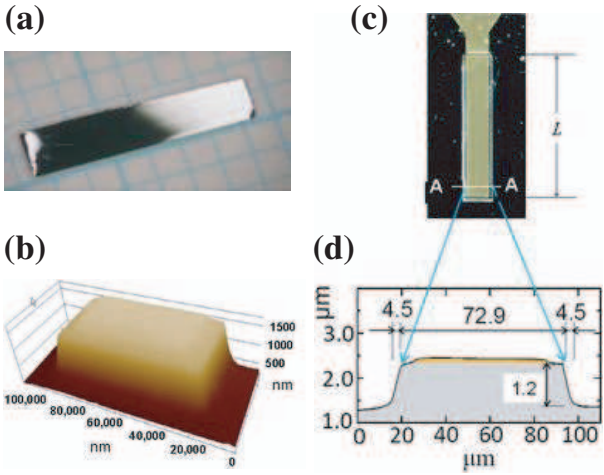}
\caption{(a): a photographs of a piece of single crystal before processing, (b): an AFM image of the mesa used in this experiment made by photolithography and argon ion milling technique, (c): the final form of the mesa used here, and (d): a view of the mesa cross-section along the A-A line.  The top and bottom widths are $w=72.9\ \mu$m, and 81.9 $\mu$m, respectively.  The top of the mesa is covered by a thin Au electrode layer.  The Au electric lead wire with 10 $\mu$m in diameter is attached by silver paste at the upper end of the gold layer shown in (c).}
\end{figure}

High-quality single crystals of BSCCO were grown by the conventional traveling solvent floating zone method using a modified infrared image furnace\cite{16,17}.  A piece of crystal was annealed overnight to an appropriate doping level at $600^\circ$C in Ar+0.1\% $\mathrm{O_2}$ atmosphere.  The resulting crystals are underdoped with $T_c\sim$ 75\ -\ 86 K.  Two types of rectangular mesas were prepared using either conventional photolithography or focused ion beam milling technique.  Photographs of a cleaved piece of a pristine single crystal, an atomic force microscope (AFM) image of a rectangular mesa after processing and the final form of the mesa are shown in Figs. 1(a), 1(b) and 1(c), respectively.  From the AFM measurements shown in Figs. 1(b) and 1(d), its length was $L\sim\ 400\ \mu$m, width $w=77.4\pm 4.5\ \mu$m and thickness $d=1.2\ \mu$m.

The temperature dependence of the normal state $c$-axis resistance $R_J(T)$ and the \textit{I-V} characteristics of the mesas were measured as described previously\cite{2,3}.  The frequency spectra of the emission were measured by a FTIR spectrometer (JASCO FARIS-1) with a Si-bolometer detector.  The angular dependence of the radiation was measured by rotating the mesa sample incrementally relative to the detector.  The solid angle of the radiation from the mesa to the detector is adjusted to have an angular resolution better than $\pm$ $2.5^{\circ}$.  The sample is cooled by a liquid $^4$He flow cryostat and the temperature is controlled to within $\pm$ 0.2 K.

The angle $\theta$ dependence of the radiation intensity ${\cal I}$ in the $xz$-plane, or $\phi=90^{\circ}$, from a similar rectangular mesa with $L\sim\ 400\ \mu$m and width $w\sim\ 60\ \mu$m are presented in Fig. 2 (a), where $\theta$ and $\phi$ are defined in Fig. 2(b).  As clearly seen in Fig. 2(a), the observed radiation is very anisotropic.  When $L/w\gg1$, as for this mesa, ${\cal I}(\theta,\phi)$ is strongest in the $xz$-plane.  The data shown in Fig. 2(a) are typical of data obtained in the $xz$-plane from several mesas.  Similar results in the $xy$-plane of the mesa are shown in Fig. 3(a), where the ${\cal I}(\theta,0^{\circ})$ data are mostly several times weaker and exhibit greater scatter than the ${\cal I}(\theta,90^{\circ})$ data.

\begin{figure}[h]
\includegraphics[width=0.5\textwidth]{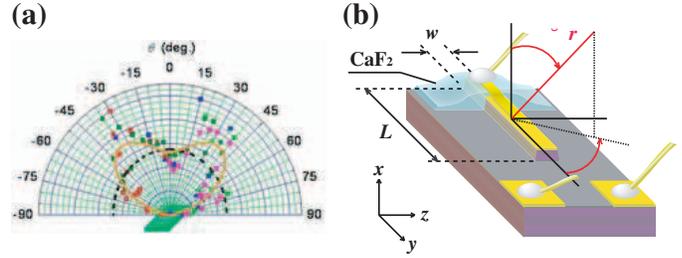}
\caption{(a): a polar plot of ${\cal I}(\theta,90^{\circ})$ normalized at ${\cal I}(0^{\circ},90^{\circ})$ measured in the $xz$-plane from a rectangular mesa at $T$=32.7 K.  Each data point was measured point by point as follows: first, cycle the \textit{I-V} curve enough slowly for only one direction to find out the radiation condition, then, stop cycling at the maximum point of radiation.  The typical values for $I$ and $V$ are 19.97 $\pm$ 0.02 mA and 0.9214 $\pm$ 0.0007 V, respectively.  The solid line presents the respective $\phi=90^{\circ}$ portion of the least-squares fits of the dual-source model to the data shown here combining the contributions from the uniform $ac$-Josephson current and the inhomogeneous displacement current exciting the (001) cavity mode, with corrections for a superconducting substrate using Model I\cite{15}.  The dashed curve presents the corresponding cavity model fit for a sample suspended in vacuum (see text). (b): a sketch of the coordinate system\cite{18}.}
\end{figure}

The general observations are summarized as follows: the maximum intensity ${\cal I}_{\rm max}$ occurs in the $xz$-plane ($\phi=90^{\circ}$) at $\theta\sim\pm 30^{\circ}$ from the mesa top ($\theta=0^{\circ}$), where ${\cal I}(0,\phi)$ is a local minimum of ${\cal I}(\theta,\phi)$, with ${\cal I}(0,90^{\circ})/{\cal I}_{\rm max}\approx$ 0.4 - 0.7, depending on the mesa measured.  This rather large sample-to-sample variation in ${\cal I}(0,90^{\circ})/{\cal I}_{\rm max}$ is not due to experimental error, but arises from specific differences in the properties of the individual mesas.  For example, the primary radiation may occur at slightly different regions in the mesa.  The thin Au electrode layer may affect the radiation.  In addition, the  spatial inhomogeneities in the $ac$-Josephson current due to sample heating and stoichiometry variations, \textit{etc.}, which are crucial for the cavity mode excitations\cite{15}, are likely to be sample-dependent.  The specific trapezoidal cross-sectional shape of each mesa as seen in Figs. 1(b) and 1(d) may also affect its angular dependence.  However, the precise reasons for this variation are not yet well understood.

If the radiation were simply induced by the fundamental cavity resonance mode, as expected from capacitor patch antenna theory\cite{18,19} and widely predicted for rectangular BSCCO mesas\cite{5,6,7,8,9,10}, ${\cal I}(\theta,\phi)$ would be maximal at $\theta=0^{\circ}$.  If a uniform $ac$-Josephson current were the primary radiation source, ${\cal I}$ would vanish at $\theta=0^{\circ}$ and be maximal near to $\theta=90^{\circ}$, as for simple dipole radiation\cite{18}.  Apparently, the experimental results contradict both simple explanations.

Second, the observed ${\cal I}(\theta,90^{\circ})$ diminishes strongly as $\theta\rightarrow 90^{\circ}$ (parallel to the $ab$-plane).  This fact contradicts the simple uniform $ac$-Josephson current model, for which a maximal intensity is expected at or near to $\theta=90^{\circ}$.  If the refractive index, $n$, were unity for the mesa as for the vacuum exterior, the radiation from a cavity mode would vanish at $\theta=90^{\circ}$, as observed.  However, for the appropriate value $n=4.2$, the predicted ${\cal I}(90^{\circ},90^{\circ})$ from the simple cavity resonance model is $\sim87$\% of its maximum value\cite{15}, as shown in Fig. 2(a), unlike the experiments.

Third, radiation, with an order of magnitude weaker than the fundamental, at several integral higher harmonics of the fundamental frequency is generally observed, as the previous report\cite{3}, but subharmonics were never observed.  Fourth, small but clear minor lobes are also observed at $\theta\sim \pm\ 75^\circ$ as seen in Figs. 2(a) and 3(a), although their integrated intensity appears to be only a few \% or less of the total.  These lobes may arise from higher harmonics, but the spectrum of the entire ${\cal I}(\theta,\phi)$ has not yet been measured.

\begin{figure}
\includegraphics[width=0.48\textwidth, trim=0 4cm 0 2cm]{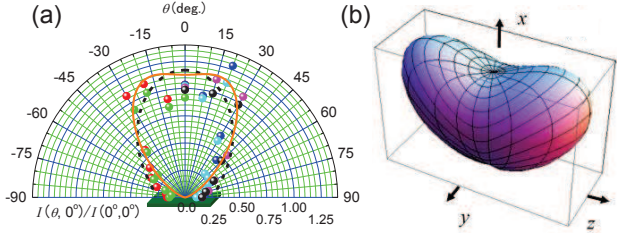}
\caption{(a): a polar plot of the observed ${\cal I}(\theta,0^{\circ})$ normalized at ${\cal I}(0^{\circ},0^{\circ})$ in the $xy$-plane of the rectangular mesa at $T$=30.0 K.  Each data point was measured in a similar manner as that shown in Fig. 2(a).  The typical values for $I$ and $V$ in this particular case are 14.90 $\pm$ 0.02 mA and 0.7995 $\pm$ 0.0005 V, respectively.  The solid curve presents the $\phi=0^{\circ}$ portion of the overall dual-source fit using Model I\cite{15}.  The dashed curve is the corresponding cavity model fit. (b): the 3D plot of the predicted ${\cal I}(\theta,\phi)$ from the least-squares fit.}
\end{figure}

These results can naturally be understood by combining both radiation sources, since there actually exist both the displacement current and the real $ac$-Josephson current flowing across the mesa, which have both spatially uniform and non-uniform parts.  Problems in fitting the data near to $\theta=\pm90^{\circ}$ can be overcome by an appropriate model of the superconducting substrate.\cite{15}

It was shown previously that the fundamental frequency $\nu$ satisfies the $ac$-Josephson relation\cite{2,3}, $\nu=\frac{2e}{h}V/N$, where $e$ is the electronic charge and $h$ is the Planck's constant.  Since $w$ is comparable to the wavelength of the plasma waves in the mesa, cavity resonances may occur as standing waves.  Empirically, the wavelength $\lambda=2w$ of the lowest energy state in the mesa, leading to the condition $\nu=c_0/n\lambda =c_0/2nw$, where $c_0$ is the speed of light in vacuum, $n=\sqrt{\epsilon}$, and $\epsilon$ is the dielectric constant of the junctions.  This empirical relation appears to work very well in many samples with different widths and $L/w$ ratios, as shown in Fig. 4.  From the slope of the line in Fig. 4, $n=4.2$ is obtained fairly accurately, corresponding to the THz-frequency dielectric constant $\epsilon=17.6$, which is about 50\% larger than the value ($\epsilon=12$) obtained by infrared spectroscopy\cite{20}.  No anomaly in $\epsilon$ for frequencies up to $0.9$ THz has been observed.

\begin{figure}
\includegraphics[width=0.45\textwidth, trim=0 1cm 2cm 0]{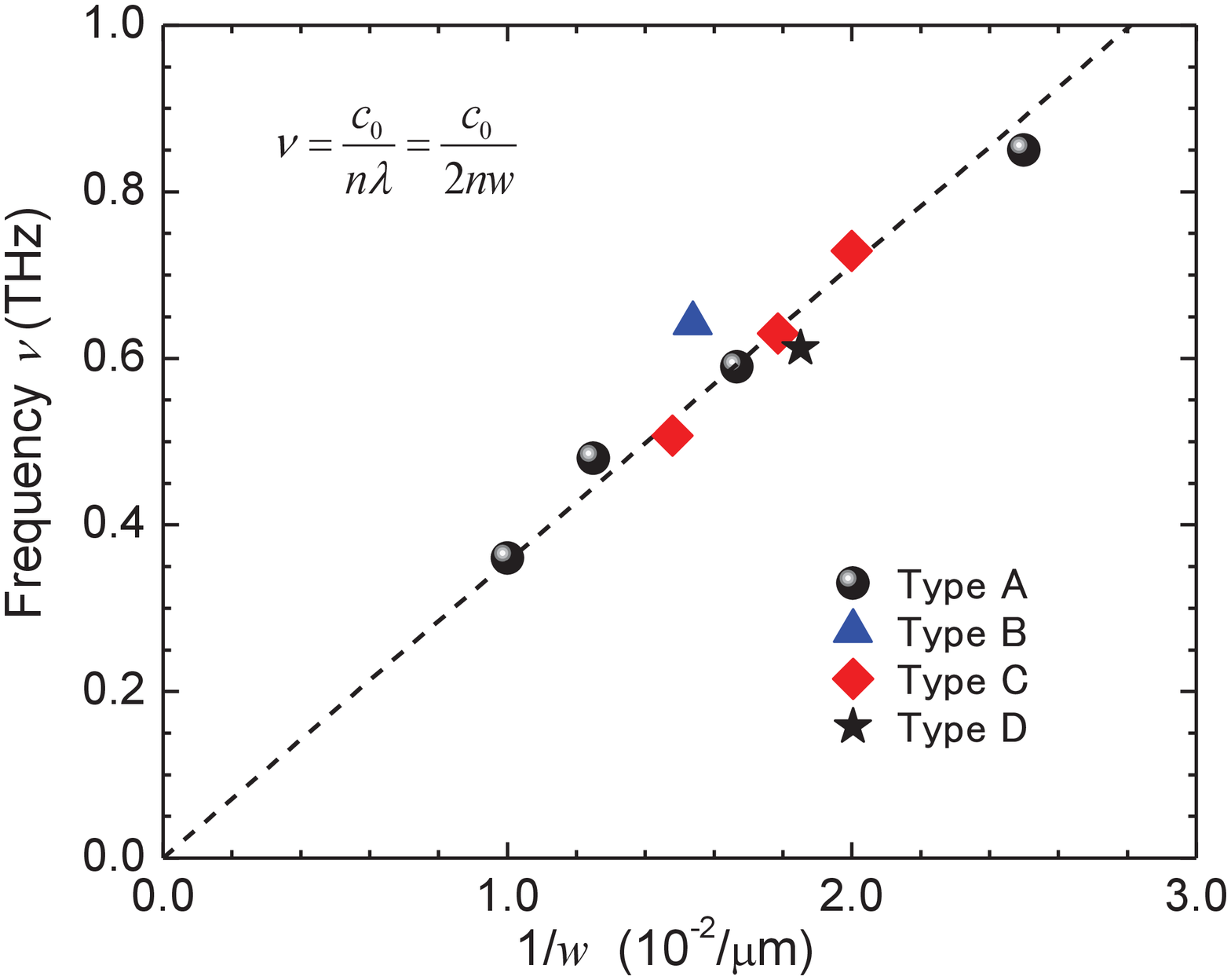}
\caption{A plot of the fundamental frequency versus $1/w$.  The data were obtained from samples with 40 $\mu$m$\leqq w\leqq100\ \mu$m and 300 $\mu$m$\leqq L\leqq400 \mu$m and variable $L/w$ ratios, prepared using four different methods from nine different single crystals.  Type A represents the data from the reference\cite{2}.  Type B, C and D were prepared by FIB, ion milling using metal mask, and photolithography techniques, respectively.}
\end{figure}

For $d\ll L, w$, the lowest mode frequencies satisfy $\nu_{0mp}=\frac{c_0}{2\sqrt{\epsilon}}\sqrt{(m/L)^2+(p/w)^2}$, where $m$ and $p$ are integers\cite{18}.  Hence, the lowest frequency mode for $L > w$ is expected to be the (010) mode with $\nu_{010}$=$\frac{c_0}{2L\sqrt{\epsilon}}$.  As seen in Fig. 4, the experimental results are definitely consistent with the (001) mode, not the (010) mode, contradicting the simple cavity resonance model.  This may be related to an instability that could arise from the energy loss by the penetration of the magnetic field in the $z$-direction into the mesa, since $\lambda\sim\lambda_c$, where $\lambda_c$ is the superconducting penetration depth in the $c$-axis direction, resulting in higher inductive energy states for the $(0m0)$ modes.  But whatever the actual reasons for the apparent non-resonant $(0m0)$ modes are, the angular dependence of the far-field radiation is calculated by combining the output from the two sources, one of which excites the (001) mode taking into account the superconducting substrate\cite{15}.  Least-squares fits to the data shown in Figs. 2(a) and 3(a) were performed using Model I, which averages the two source contributions symmetrically about $\theta=0^{\circ}$, while preserving the Love magnetic equivalence principle boundary condition $\hat{\bm n}\times{\bm H}=0$ on the mesa edge\cite{15,18}, and using Model II, which relaxes the boundary condition while preserving $\lambda$\cite{15}.  The best fit with standard deviation $\sigma=0.122$ was found for Model I with mixing parameter $\alpha_1(0)=0.310$, corresponding to 24\% of the overall intensity arising from the magnetic (cavity) source.  The cross-sections of the best dual- and single-source fits are shown by the solid- and dashed- curves in Figs. 2(a) and 3(a), and the full radiation pattern resulting from the dual-source best fit is shown in Fig. 3(b).  Preliminary but similar results were obtained from numerical simulations\cite{11,12,14}.

Figure 5 presents the detailed $dc$ \textit{I-V} characteristics and the corresponding output of the Si-bolometer detector as functions of $V$, when the scan is made very slowly (2.8 hours for a one-cycle measurement).  In this case many radiation peaks are observed.  Examining them in detail, one finds an interesting feature.  When $I$ is reduced slowly on the return \textit{I-V} branch, the radiation energy starts to build up gradually until the growth is interrupted by a jump to a different \textit{I-V} branch.  This is more clearly seen in the inset of Fig. 5, in which the sawtooth-like radiation peaks are seen over the range of $V$ from -0.67 to -0.57 V.  From the detailed FTIR measurements performed at various points in the sawtooth region, it turns out that the observed frequency shifts to lower frequency as the power increases (\textit{i.e.}, $I$ decreases).  This can be interpreted as the gradual decrease of the $ac$-Josephson frequency as $I$ decreases, until the mesa eventually enters a cavity resonance state.  Meanwhile, the radiation energy increases in the mesa, and when the power reaches a threshold, $I$ jumps to a different \textit{I-V} branch.  Therefore, the mechanism of the strong coherent continuous radiation appears to be due to gradual development of the synchronization in the individual or groups of Josephson junctions onto the cavity resonance mode, rather than a sudden catastrophic transition.  This fact strongly suggests that the stronger output power may be achieved by preventing the \textit{I-V} curve from jumping.

\begin{figure}
\includegraphics[width=0.5\textwidth, trim=1cm 1.5cm 0 0]{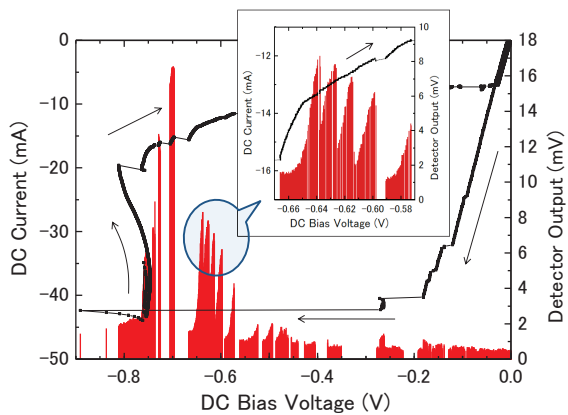}
\caption{The \textit{I-V} characteristic (connected data-point representation) and the radiation power (vertical line representation) detected by the Si-bolometer at $T$=30.0 K.  The arrows indicate the sweeping direction of the \textit{I-V} curve.  The detector was set at the angle $\theta$=45$^{\circ}$ and $\phi$=90$^{\circ}$ for this particular measurement.  The inset is a magnified view of the encircled region.}
\end{figure}

The combination of two conditions necessary for the resonant emission may be written as $V_{dc}^{tot}$=$(c_0/2nw)(1/K_J)(2d/c)$, where $V_{dc}^{tot}$ is the required $dc$ voltage to meet the cavity resonance oscillation, the Josephson constant $K_J$=483.5979 (GHz/mV), and $c\approx$ 30.65 \AA{\hskip1pt} is the $c$-axis lattice constant of BSCCO.  This reduces to $V_{dc}^{tot}$=$48.2(d/w)$, which surprisingly depends only upon $d/w$\cite{21}.  Since an emitting junction must be in its resistive state, its resistance $R_J$ and $dc$ current $I$ must satisfy $R_JI\ge V_{dc}^{tot}$.  Because $R_J(T)$ has a steep negative temperature coefficient, and BSCCO each junction is subject to severe Joule heating, the whole mesa may thermally be stabilized.  On the other hand, because of large reduction of $R_J$ the above condition for synchronized emission often may not be satisfied, especially for overdoped samples, due to their considerably smaller $R_J$ values.  This may explain partially why we initially had many unsuccessful mesas.

In conclusion, we measured the angular dependence of the far-field THz emission from mesas of single-crystal $\mathrm{Bi_2Sr_2CaCu_2O_{8+\delta}}$.  Our results provide strong evidence for a dual-source mechanism in which the uniform and inhomogeneous parts of the $ac$-Josephson current respectively act as an electric surface current source and set up a displacement current that excites a mesa cavity resonance mode which locks the radiation frequency, and acts as a magnetic surface current source.  By adjusting the relative amplitude and phase of the two source currents and accounting for the substrate, excellent agreement with experimental results is obtained.  From FTIR analysis of the \textit{I-V} characteristic, the radiation appears to build-up from gradual synchronization of the individual or groups of Josephson junctions to the cavity resonance mode.

The authors are deeply indebted to Drs. U. Welp, A. E. Koshelev and W.-K. Kwok for close collaborations.  We thank Prof. N. Pedersen and Dr. X. Hu and Mr. S. Lin for stimulating and fruitful discussions.  This work was supported in part by CREST-JST (Japan Science and Technology Agency), WPI (World Premier International Research Center Initiative)-MANA (Materials Nanoarchitechtonics) project (NIMS) and Strategic Initiative category (A) at the University of Tsukuba.

\end{document}